\documentclass[pdftex,preprint]{svjour3}

\usepackage[dvips]{graphicx}
\usepackage{amssymb}
\usepackage{amsbsy}
\usepackage{pslatex}

\newcommand{\bkp}{\boldsymbol{\kappa}}

\begin{document}

\title{Strongly correlated electron system in the magnetic field}

\author{A. Sherman \and M. Schreiber}

\institute{A. Sherman \at Institute of Physics, University of Tartu, Riia 142,
51014 Tartu, Estonia\\
              Tel.: +372-7-374616\\
              Fax: +372-7-383033\\
              \email{alexei@fi.tartu.ee}
              \and M. Schreiber \at Institut f\"ur Physik, Technische Universit\"at Chemnitz, D-09107 Chemnitz, Germany}

\date{Received: date / Accepted: date}

\maketitle

\begin{abstract}
The energy spectrum of the two-dimensional $t$-$J$ model in a perpendicular magnetic field is investigated. The density of states at the Fermi level as a function of the inverse magnetic field $\frac{1}{B}$ reveals oscillations in the range of hole concentrations $0.08<x<0.18$. In the used approximation zero-field Fermi surfaces are large for these $x$, and oscillation frequencies conform with such Fermi surfaces. However, the amplitude of these oscillations is modulated with a frequency which is smaller by an order of magnitude. The appearance of this modulation is related to van Hove singularities in the Landau subbands, which traverse the Fermi level with changing $B$. The singularities are connected with bending the Landau subbands due to strong electron correlations. The frequency of the modulation is of the same order of magnitude as the quantum oscillation frequency observed in underdoped cuprates.
\end{abstract}

\section{Introduction}
Theoretical investigations of systems of strongly correlated electrons in strong magnetic fields were started shortly after the discovery of the high-$T_c$ superconductivity. A number of works was carried out on small clusters using the exact diagonalization (see, e.g., \cite{Castello,Beran,Albuquerque}). It is worth noting that due to the Peierls factor \cite{Peierls}, the translation symmetry of the system is changed \cite{Brown} -- in the magnetic field the size of the elementary cell increases significantly. Clusters with sizes smaller than the size of this supercell violate the symmetry of the Hamiltonian and, therefore, it is difficult to extend the obtained results to larger crystals. Another approximation used for this problem is the mean-field approximation (see, e.g., \cite{Tripathi,Yang,Schmid}). The main shortcoming of this approximation is the neglect of the dynamic interaction of fermions with spin excitations. This interaction defines the fermion dispersion in the underdoped case.

The investigation of a system with strong electron correlations in the magnetic field is of interest in connection with the observation of quantum oscillations in the mixed state of underdoped yttrium cuprates \cite{Doiron,Bangura,Yelland,Sebastian08}. Based on the Onsager-Lifshitz-Kosevich theory for metals \cite{Shoenberg} the observed decreased quantum oscillation frequencies were interpreted as a manifestation of small Fermi surface pockets \cite{Sebastian12}, despite the fact that this interpretation seems to be in contradiction with numerous photoemission experiments. To explain the appearance of these small pockets proposals for various states with broken translational symmetry were suggested \cite{Millis,Chen,Galitski}. Other theories for explaining the decreased quantum oscillation frequency suppose that it is connected with superconducting fluctuations \cite{Melikyan,Pereg} or use phenomenology of the marginal Fermi liquid \cite{Varma}.

Crystals, in which the decreased quantum oscillation frequencies were observed, belong to the underdoped region of the cuprate phase diagram, and, therefore, they are characterized by strong electron correlations. Theoretically the behavior of such crystals in strong magnetic fields is poorly known. In this work we use an approach, which allows us to overcome some limitations of the approximations used earlier for this problem. The approach provides a way to consider large enough clusters and moderate magnetic fields, taking into account the interaction of holes with spin excitations. We use the exact diagonalization of the kinetic term of the two-dimensional (2D) $t$-$J$ model of Cu-O planes. The term contains the Peierls factor describing the influence of the homogeneous magnetic field perpendicular to the plane. As known \cite{Brown,Langbein,Hsu}, the energy spectrum of the kinetic term consists of the Landau subbands, which substitute the Landau levels in the lattice problem. The influence of strong electron correlations on these subbands is investigated using the Mori projection operator technique \cite{Mori}. It should be noted that due to the complexity of the problem we were able to calculate only a few terms of the continued fraction of Green's function. Therefore, the obtained zero-field normal-state spectrum does not contain the pseudogap, which appears in other approaches both in the $t$-$J$ and Hubbard models \cite{Sherman97,Sherman06,Macridin}. Since in the pseudogap the density of states (DOS) does not completely vanish, we nevertheless suppose that the used approximation gives at least qualitatively a correct picture of the normal-state energy spectrum of the $t$-$J$ model in the perpendicular magnetic field.

In the used approximation, in the normal state the model has a large Fermi surface for the hole concentrations $x\gtrsim 0.06$. We found that in the range $0.08<x<0.18$ the density of hole states oscillates with frequency for a fixed field $B$ and with $\frac{1}{B}$ at the Fermi level. The frequency of these latter oscillations with $\frac{1}{B}$ conforms with the large Fermi surface. However, the amplitude of these high-frequency oscillations is modulated with a frequency which is smaller by an order of magnitude. The appearance of this modulation is related to van Hove singularities in the Landau subbands, which traverse the Fermi level with changing $B$. These van Hove singularities are connected with bending the Landau subbands due to the influence of strong correlations. The frequency of the modulation is of the same order of magnitude as the quantum oscillation frequency observed in underdoped cuprates.

\section{Main formulas}
One of the main models used for the description of Cu-O planes of cuprate superconductors is the 2D $t$-$J$ model. The Hamiltonian of this model in the magnetic field, which is perpendicular to the plane, reads
\begin{eqnarray}\label{hamiltonian}
H&=&\sum_{\bf ll'\sigma}t_{\bf ll'}\exp\left(i\frac{e}{\hbar}\int_{\bf l}^{\bf l'}\!\!\!{\bf A}({\bf r})\,d{\bf r}\right)a^\dagger_{\bf l\sigma}a_{\bf l'\sigma}\nonumber\\
&&+\frac{1}{2}\sum_{\bf ll'}J_{\bf ll'}\left(s^z_{\bf l}s^z_{\bf l'}+s^+_{\bf l}s^-_{\bf l'}\right)+g\mu_BB\sum_{\bf l}s^z_{\bf l},
\end{eqnarray}
where 2D vectors ${\bf l}$ and ${\bf l'}$ label sites of a square plane lattice, $\sigma=\pm 1$ is the projection of the hole spin, $a^\dagger_{\bf l\sigma}=|{\bf l}0\rangle\langle{\bf l}\sigma|$ and $a_{\bf l\sigma}=|{\bf l}\sigma\rangle\langle{\bf l}0|$ are hole creation and annihilation operators with the empty $|{\bf l}0\rangle$ and singly occupied $|{\bf l}\sigma\rangle$ site states. These three states form the complete set of hole states for the site {\bf l} in the $t$-$J$ model. The first term of the Hamiltonian, the hole kinetic energy $H_k$, contains the hopping matrix element $t_{\bf ll'}$ and the exponential factor with the Peierls phase \cite{Peierls}, in which ${\bf A}({\bf r})$ is the vector potential. The second term on the right-hand side of (\ref{hamiltonian}) is the exchange energy of localized spins with the exchange constant $J_{\bf ll'}$ and the spin-$\frac{1}{2}$ operators $s^z_{\bf l}=\frac{1}{2}\sum_\sigma \sigma|{\bf l}\sigma\rangle\langle{\bf l}\sigma|$ and $s^\pm_{\bf l}=|{\bf l},\pm 1\rangle\langle{\bf l},\mp 1|$. The last, Zeeman term of the Hamiltonian contains the g-factor $g\approx 2$, the Bohr magneton $\mu_B$ and the magnetic induction $B$ of the external magnetic field. It is supposed that this field is homogeneous and is only weakly disturbed by internal currents \cite{Atkinson}.

In the following consideration we shall suppose that only the nearest neighbor hopping and exchange constants are nonzero,
$$t_{\bf ll'}=t\sum_{\bf a}\delta_{\bf l,l'+a},\quad J_{\bf ll'}=J\sum_{\bf a}\delta_{\bf l,l'+a},$$
where ${\bf a}$ are four vectors connecting nearest neighbor sites. In cuprate perovskites, the exchange constant $J$ is of the order of 100~meV. Comparing the interaction energy between a spin with its four neighbors (the second term of the Hamiltonian) and the energy of the spin in the external field (the third term of the Hamiltonian) one can ascertain that the former energy is two orders of magnitude larger than the latter even for fields of the order of 50~T. Therefore, the Zeeman term of the Hamiltonian can be neglected.

In the Landau gauge ${\bf A_l}=-Bl_y{\bf x}$, where $l_y$ is the $y$ component of the site vector ${\bf l}$ and ${\bf x}$ is the unit vector along the $x$ axis (the 2D lattice is located in the $xy$ plane, and the field is directed along the $z$ axis). Hence the exponential in the kinetic term of the Hamiltonian can be written as
\begin{equation}\label{pphase}
{\rm e}^{i\bkp_{\bf a}{\bf l}},\quad \bkp_{\bf a}=-\frac{e}{\hbar}Ba_x{\bf y}.
\end{equation}

In the following discussion we shall restrict our consideration to the fields satisfying the condition
\begin{equation}\label{condition}
\frac{e}{\hbar}Ba^2=\frac{2\pi}{n},
\end{equation}
where $a=|{\bf a}|$ and $n$ is an integer. In this case the kinetic term of the Hamiltonian defines its translation properties -- $H_k$ is invariant with respect to translations by the lattice period along the $x$ axis and by $n$ lattice periods along the $y$ axis. To retain this symmetry we apply the periodic Born-von Karman boundary conditions to the sample with $N_x$ sites along the $x$ axis and $nN_y$ sites along the $y$ axis. The boundary conditions define the set of allowed wave vectors with components $K_x=\frac{2\pi}{N_xa}n_x$ and $K_y=\frac{2\pi}{nN_ya}n_y$ with integer $n_x$ and $n_y$. As can be seen from (\ref{pphase}) and (\ref{condition}), the momenta $\bkp_{\bf a}$ coincide with one of the wave vectors in this net. Therefore, in $H_k$ we can perform the usual Fourier transformation
$$a_{\bf l\sigma}=\frac{1}{\sqrt{N}}\sum_{\bf K}{\rm e}^{-i{\bf Kl}}a_{\bf K\sigma},$$
using the known result $\frac{1}{N}\sum_{\bf l}{\rm e}^{i{\bf Kl}}=\delta_{{\bf K,Q}}$ for the wave vector ${\bf K}$ on the net. Here $N=nN_xN_y$ and ${\bf Q}=\left(\frac{2\pi}{a}\nu_x,\frac{2\pi}{a} \nu_y\right)$ with integer $\nu_x$ and $\nu_y$. After the Fourier transformation the kinetic term acquires the form
\begin{equation}\label{kterm}
H_k=t\sum_{\bf Ka\sigma}{\rm e}^{i{\bf Ka}}a^\dagger_{{\bf K}-\bkp_{\bf a},\sigma}a_{\bf K\sigma}.
\end{equation}
In deriving (\ref{kterm}) we took into account that $\bkp_{\bf a}{\bf a}=0$.

It is convenient to split the Brillouin zone into $n$ stripes of the width $\frac{2\pi}{na}$, which are oriented parallel to the $x$ axis. If we select one of these stripes, say, the lowest one with $-\frac{\pi}{a}<K_y\leq -\frac{\pi}{a} +\frac{2\pi}{na}$, and denote wave vectors in it as ${\bf k}$,  momenta in the entire Brillouin zone can be described as ${\bf k}+j\bkp$. Here $0\leq j \leq n-1$ and $\bkp=\frac{2\pi}{na}{\bf y}$. In these notations the kinetic energy acquires the form
\begin{equation}\label{kmatr}
H_k=\sum_{\bf k\sigma}{\bf A^\dagger_{k\sigma}h_k A_{k\sigma}},
\end{equation}
where the summation over ${\bf k}$ is performed over the selected stripe,
$${\bf A^\dagger_{\bf k\sigma}}=\left(a^\dagger_{\bf k\sigma},a^\dagger_{\bf k+\bkp,\sigma},\ldots a^\dagger_{{\bf k}+(n-1)\bkp,\sigma}\right),$$
\begin{equation}\label{hmatr}
{\bf h_k}=t\left(\begin{array}{cccccc}
q_0 & {\rm e}^{-ik_x a} & 0 & 0 & \ldots & {\rm e}^{ik_x a} \\
{\rm e}^{ik_x a} & q_1 & {\rm e}^{-ik_x a} & 0 & \ldots & 0 \\
0 & {\rm e}^{ik_x a} & q_2 & {\rm e}^{-ik_x a} & \ldots & 0 \\
\vdots & \vdots & \vdots & \vdots & \vdots & \vdots \\
{\rm e}^{-ik_x a} & 0 & \ldots & 0 & {\rm e}^{ik_x a} & q_{n-1}
\end{array}\right),
\end{equation}
and $q_j=2\cos\left(k_y a+\frac{2j\pi}{n}\right)$.

The cyclic Hermitian matrix (\ref{hmatr}) can be diagonalized by the unitary transformation
\begin{eqnarray}
&&a_{{\bf k}+j\bkp,\sigma}=\sum_{m=0}^{n-1} U_{{\bf k}jm}\alpha_{{\bf k}m\sigma}\nonumber\\[-1ex]
&&\label{transf}\\[-1ex]
&&\sum_{j'} h_{{\bf k}jj'}U_{{\bf k}j'm}=E_{{\bf k}m}U_{{\bf k}jm}. \nonumber
\end{eqnarray}
This diagonalization gives the dispersion of the Landau subbands $E_{{\bf k}m}$ of uncorrelated carriers in the reduced Brillouin zone, which coincides with the above-mentioned lowest stripe. The applied procedure for deriving this dispersion is equivalent to the usually used approach, in which the magnetic supercell is introduced (see, e.g., \cite{Brown,Atkinson}). The derivation of the kinetic-energy matrix (\ref{hmatr}) is simpler in our approach.

Since the kinetic energy defines symmetry properties of the total Hamiltonian (\ref{hamiltonian}), states corresponding to operators $\alpha_{{\bf k}m\sigma}$ in (\ref{transf}) are appropriate starting states for the consideration of the influence of spin excitations on hole states. For this purpose we calculate Green's function
\begin{equation}\label{green}
G({\bf k}m\bar{t})=-i\theta(\bar{t})\left\langle\left\{\alpha^\dagger_{{\bf k}m\sigma}(\bar{t}),\alpha_{{\bf k}m\sigma}\right\}\right\rangle,
\end{equation}
where the averaging over the grand canonical ensemble and the operator time dependence are determined by the Hamiltonian ${\cal H}=H-\mu \sum_{\bf l}|{\bf l}0\rangle\langle{\bf l}0|$,
\begin{eqnarray*}
&&\langle\ldots\rangle={\rm Tr} \left({\rm e}^{-\beta{\cal H}}\ldots\right)/{\rm Tr} \left({\rm e}^{-\beta{\cal H}}\right),\\ &&\alpha^\dagger_{{\bf k}m\sigma}(\bar{t})={\rm e}^{i{\cal H}\bar{t}}\alpha^\dagger_{{\bf k}m\sigma}{\rm e}^{-i{\cal H}\bar{t}},
\end{eqnarray*}
$\beta$ and $\mu$ are the inverse temperature and the chemical potential, respectively. Green's function (\ref{green}) does not depend on the spin projection.

To calculate this function we use the Mori projection operator technique \cite{Mori}. In this approach, the Fourier transform of Green's function (\ref{green}) is represented by the continued fraction
\begin{equation}\label{cfraction}
G({\bf k}m\omega)=\frac{\left\langle\left\{\alpha_{{\bf k}m\sigma}, \alpha^\dagger_{{\bf k}m\sigma}\right\}\right\rangle
}{\omega-E_0-\frac{\textstyle V_0}{\textstyle\omega-E_1-\frac{\textstyle V_1}{\ddots}}},
\end{equation}
where the elements of the fraction are obtained from the recurrence relations \cite{Sherman87,Sherman02}
\begin{eqnarray}
&&A_{k+1}=[A_k,{\cal H}]-V_{k-1}A_{k-1}-E_k A_k,\nonumber\\
&&E_k=\left\langle\left\{[A_k,{\cal H}],A_k^\dagger\right\}\right\rangle
 \left\langle\left\{A_k,A_k^\dagger\right\}\right\rangle^{-1},
 \nonumber\\[-1.5ex]
&&\label{rrelations}\\[-1.5ex]
&&V_k=\left\langle\left\{ A_{k+1},A_{k+1}^\dagger\right\}\right\rangle \left\langle\left\{A_k,A_k^\dagger\right\}\right\rangle^{-1},\nonumber\\
&&V_{-1}=0,\; A_0=\alpha_{{\bf k}m\sigma},\; k=0,1,2\ldots\nonumber
\end{eqnarray}
Operators $A_k$ depend on ${\bf k}$, $m$ and $\sigma$ and numbers $E_k$ and $V_k$ on ${\bf k}$ and $m$. To simplify notations we suppressed these dependencies in (\ref{cfraction}) and (\ref{rrelations}).

Commutation relations for the hole and spin operators, which are necessary for calculating the elements of the continued fraction (\ref{cfraction}), can be derived from their definitions in terms of bra and ket vectors of the site states. Notice that the hole operators do not satisfy the usual fermion anticommutation relations due to the prohibition of the double occupancy of sites by holes. Nevertheless,
$$\left\langle\left\{\alpha_{{\bf k}m\sigma}, \alpha^\dagger_{{\bf k'}m'\sigma'}\right\}\right\rangle \propto\delta_{\bf kk'}\delta_{mm'}\delta_{\sigma\sigma'}.$$
Therefore, we do not use the matrix form of the Mori formalism and consider only the diagonal Green's function (\ref{green}).

Using the commutation relations for the hole and spin operators we get for the elements of the continued fraction (\ref{cfraction})
\begin{eqnarray}
&&\left\langle\left\{\alpha_{{\bf k}m\sigma},\alpha^\dagger_{{\bf k}m\sigma}\right\}\right\rangle=\phi,\label{numerator}\\
&&E_0=\sum_{j=0}^{n-1}\sum_{\bf a}U^*_{{\bf k}jm}U_{{\bf k},j+j_{\bf a},m}\nonumber\\
&&\qquad\times\bigg[{\rm e}^{i({\bf k}+j\bkp){\bf a}}\bigg(t\phi\
+\frac{3tC_1}{2\phi}\bigg)+\frac{tF_1}{\phi}\bigg]\nonumber\\
&&\qquad -\frac{3JF_1}{\phi}\sum_{j=0}^{n-1}|U_{{\bf k}jm}|^2\gamma_{{\bf k}+j\bkp}-\frac{3JC_1}{\phi}-\mu,\label{e0}\\
&&V_0=\sum_{j=0}^{n-1}\sum_{\bf aa'}U^*_{{\bf k}jm}U_{{\bf k},j+j_{\bf a}-j_{\bf a'},m}{\rm e}^{i({\bf k}+j\bkp)({\bf a-a'})-i\bkp_{\bf a}{\bf a'}}\nonumber\\
&&\qquad\times t^2\bigg(\phi^2+\frac{3C_1}{2\phi}+\frac{3C_{\bf a-a'}}{2}\bigg)\nonumber\\
&&\qquad -t(\mu+E_0)\sum_{j=0}^{n-1}\sum_{\bf a}U^*_{{\bf k}jm}U_{{\bf k},j+j_{\bf a},m}\nonumber\\
&&\qquad\times\bigg[{\rm e}^{i({\bf k}+j\bkp){\bf a}}
\bigg(2\phi+\frac{3C_1}{\phi}\bigg)+\frac{2F_1}{\phi}\bigg]\nonumber\\ &&\qquad+\frac{t^2F_1}{\phi}{\rm Re}\sum_{j=0}^{n-1}\sum_{\bf aa'}U^*_{{\bf k}jm}U_{{\bf k},j+j_{\bf a}-j_{\bf a'},m}{\rm e}^{i({\bf k}+j\bkp){\bf a}}\nonumber\\
&&\qquad -\frac{t^2F_1}{\phi}\sum_{j=0}^{n-1}\sum_{\bf a}U^*_{{\bf k}jm}U_{{\bf k},j+2j_{\bf a},m}{\rm e}^{i({\bf k}+j\bkp){\bf a}}
\nonumber\\
&&\qquad +(\mu+E_0)^2+t^2x-\frac{4t^2C_1}{\phi},\label{v0}\\
&&E_1\approx -\mu. \label{e1}
\end{eqnarray}
In the above equations, $\phi=\frac{1+x}{2}$, $j_{\bf a}=-a_x/a$,
\begin{eqnarray}\label{x}
x&=&\frac{1}{N}\sum_{\bf l}\Big\langle|{\bf l}0\rangle\langle{\bf l}0|\Big\rangle\nonumber\\
&=&-\frac{1}{N\pi} \sum_{{\bf k}m}\int_{-\infty}^\infty d\omega\frac{{\rm Im}G({\bf k}m\omega)}{{\rm e}^{\beta\omega}+1}
\end{eqnarray}
is the hole concentration, $\gamma_{\bf k}=\frac{1}{4}\sum_{\bf a}{\rm e}^{i{\bf ka}}$,
\begin{eqnarray}\label{f1}
F_1&=&\frac{1}{4N}\sum_{\bf la}\left\langle a^\dagger_{\bf l\sigma} a_{\bf l+a,\sigma}\right\rangle\nonumber\\
&=&\frac{1}{N}\sum_{{\bf k}jm}\gamma_{{\bf k}+j\bkp}|U_{{\bf k}jm}|^2 \int_{-\infty}^\infty d\omega\frac{{\rm Im}G({\bf k}m\omega)}{{\rm e}^{\beta\omega}+1},
\end{eqnarray}
$C_1=\frac{1}{4N}\sum_{\bf la}\left\langle s^+_{\bf l}s^-_{\bf l+a} \right\rangle$ and $C_{\bf a-a'}=\frac{1}{N}\sum_{\bf l}\left\langle s^+_{\bf l}s^-_{\bf l+a-a'}\right\rangle$.

In equations (\ref{e0})--(\ref{e1}), some terms were obtained by decoupling of operators belonging to different lattice sites. The quantity $E_1$ is described by a more complex expression than that given by (\ref{e1}). The comparison of results obtained with this more complex $E_1$ and with that in (\ref{e1}) shows their similarity. Therefore, to speed up calculations we used the latter value for $E_1$. Quantities $\big\langle |{\bf l}0\rangle\langle{\bf l}0|\big\rangle$ and $\big\langle a^\dagger_{\bf l\sigma} a_{\bf l+a,\sigma}\big\rangle$, which actually appear in (\ref{e0}) and (\ref{v0}), vary slightly within the magnetic supercell (in the considered case it is a row of $n$ sites along the $y$ direction). However, the variation is small -- for example, for $x=0.14$ and $n=64$ the maximum relative deviation of $\big\langle|{\bf l}0\rangle\langle{\bf l}0|\big\rangle$ from $x$ is less than 0.02. Therefore, in the above expressions we substituted these quantities by their averaged over the crystal values $x$ and $F_1$. We supposed that parameters $C_1$ and $C_{\bf a-a'}$, which characterize the spin subsystem, are approximately the same as for $B=0$. We took these latter parameters from the results of self-consistent calculations carried out for the case of zero temperature \cite{Sherman04},
\begin{eqnarray}
&&C_1=-\frac{0.207}{1+3.12x+1.17x^2},\nonumber\\
&&C_{\bf a-a'}=\frac{1}{N}\sqrt{\frac{|C_1|}{\alpha}}\sum_{\bf K} \cos[{\bf K}({\bf a-a'})]\nonumber\\[-1.5ex]
&&\label{spins}\\[-1.5ex]
&&\qquad\qquad\times\sqrt{\frac{1-\gamma_{\bf K}}{\Delta+1+ \gamma_{\bf K}}},\nonumber\\
&&\Delta=2x^2,\quad\alpha=1.802-0.802\tanh(10x),\nonumber
\end{eqnarray}
where the summation is performed over the entire Brillouin zone.

The calculation of the elements of the continued fraction (\ref{cfraction}) is terminated at this stage. As a consequence for all values of ${\bf k}$ in the reduced Brillouin zone and $m$ in the range $[0,n-1]$ we get Green's function with two poles,
\begin{eqnarray}
&&G({\bf k}m\omega)=\frac{\phi(E_1-\omega)}{\varepsilon_+ -\varepsilon_-}\left(\frac{1}{\omega-\varepsilon_-}-
\frac{1}{\omega-\varepsilon_-}\right),\nonumber\\[-1.5ex]
&&\label{twopoles}\\[-1.5ex]
&&\varepsilon_\pm=\frac{E_1+E_0}{2}\pm\sqrt{\frac{(E_1-E_0)^2}{4}+ V_0}.\nonumber
\end{eqnarray}
As can be seen from equations~(\ref{numerator})--(\ref{v0}), Green's function~(\ref{twopoles}) depends on parameters $x$ and $F_1$, which, in their turn, are calculated from this function using equations~(\ref{x}) and (\ref{f1}). These parameters were calculated self-consistently.

\begin{figure}[htb]
\resizebox{\textwidth}{!}{\includegraphics[width=\textwidth]{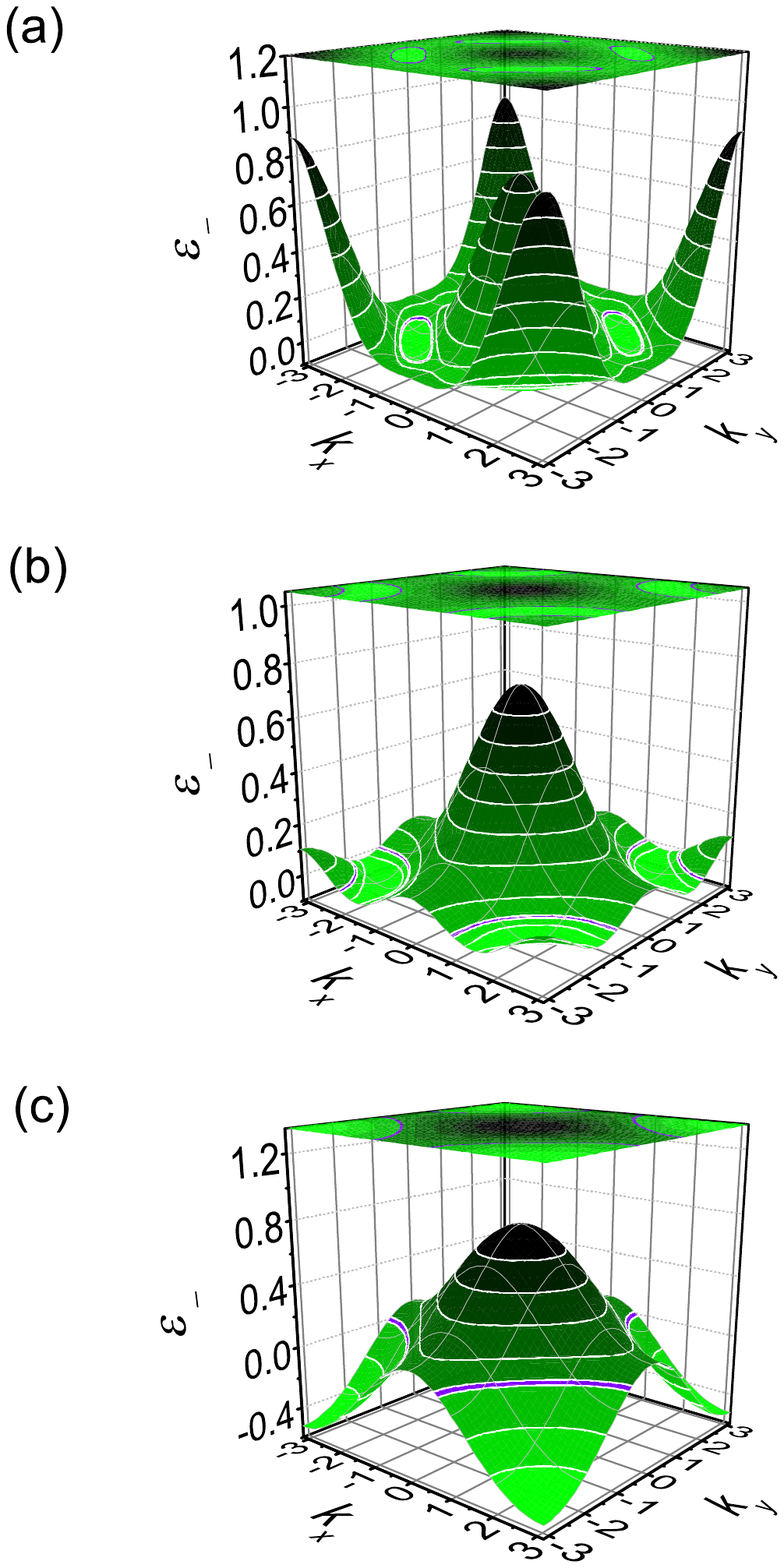}}
\caption{The hole dispersion in the case $B=0$ for the hole concentrations $x=0.033$ (a), $0.076$ (b) and $0.14$ (c). $J/t=0.2$, $T=0$. The Fermi surface is shown by the violet (dark) contour. } \label{Fig1}
\end{figure}
In \cite{Sherman04} this approach was used for investigating the evolution of the hole dispersion with doping in the 2D $t$-$J$ model in the case $B=0$. Equations of that work can be obtained from (\ref{e0})--(\ref{f1}) by setting $\bkp=\bkp_{\bf a}=0$, $j=j_{\bf a}=0$, $U_{{\bf k}jm}= \delta_{j0}\delta_{m0}$ and ${\bf k\rightarrow K}$. It was shown that the pole with the lower energy $\varepsilon_-$ corresponds to the spin-polaron band, which gives the most intensive maximum in the hole spectral function of the 2D $t$-$J$ model. The pole with the higher energy $\varepsilon_+$ imitates other excitations. As known from results obtained by other methods, these higher-energy excitations are the electron-spin excitation continuum and low intensity string states (see the reviews \cite{Izyumov,Dagotto} on the $t$-$J$ model). We are mainly interested in states near the Fermi level, where the spin-polaron states dominate at low hole concentrations. The dispersion of this band is shown in Fig.~\ref{Fig1} for several values of $x$.
\begin{figure}[htb]
\resizebox{\textwidth}{!}{\includegraphics[width=\textwidth]{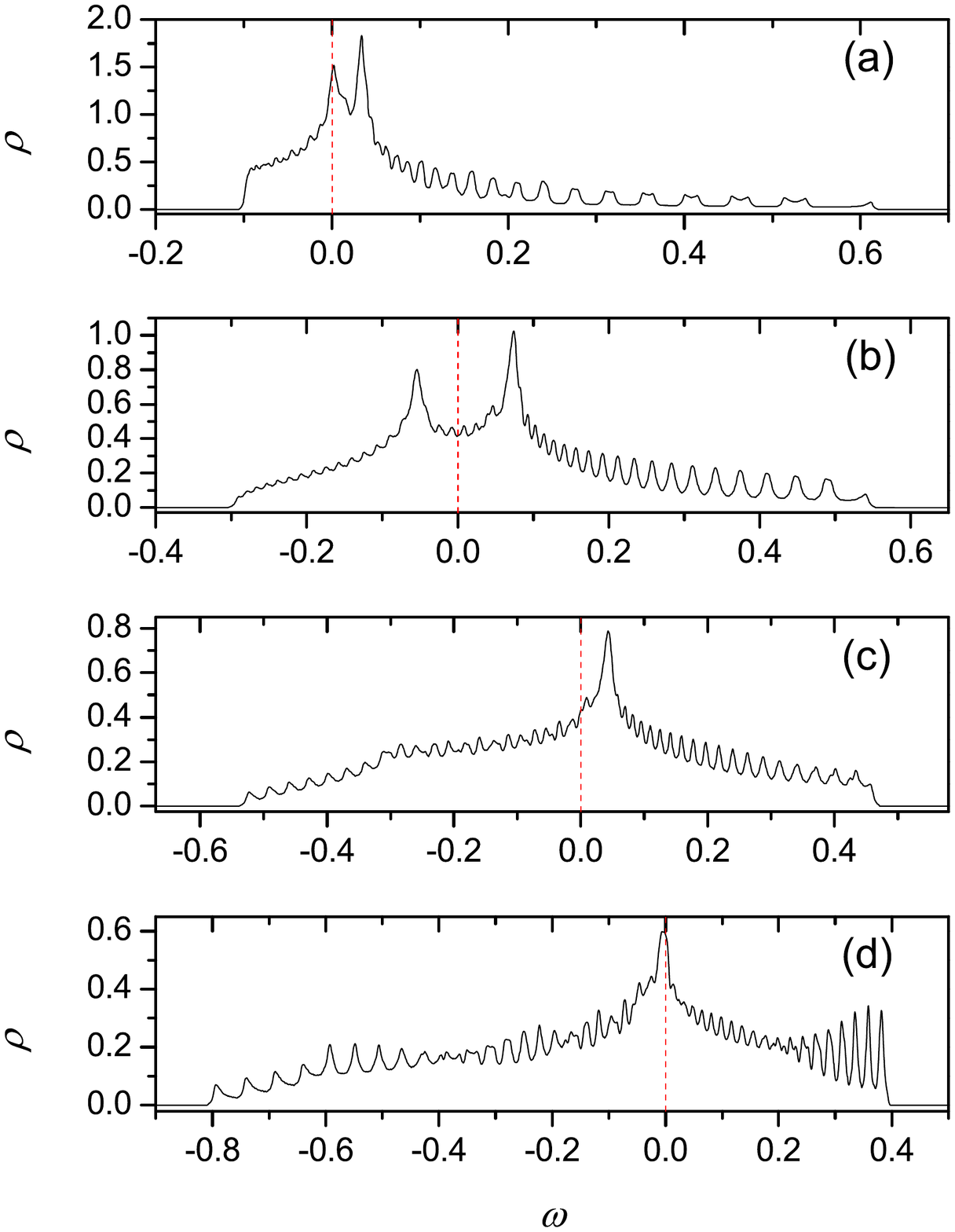}}
\caption{The density of hole states in the case $n=47$ for the hole concentrations $x=0.076$ (a), $0.11$ (b), $0.14$ (c) and $0.18$ (d). The Fermi level is shown by the red dashed line.} \label{Fig2}
\end{figure}
Here and hereafter we set $J/t=0.2$, the temperature $T=0$, energies and lengths are measured in units of $t$ and $a$, respectively. The evolution of the hole dispersion with doping, which is presented in Fig.~\ref{Fig1}, can be understood in the following way. In the heavily underdoped case $x=0.033$ the crystal is characterized by the short-range antiferromagnetic ordering, which is reflected in the gap in the spin-excitation spectrum at the antiferromagnetic momentum $(\pi,\pi)$. The gap is described by the parameter $\Delta$, equation~(\ref{spins}), which is connected with the correlation length of the short-range order $\xi$ by the relation $\xi\approx\frac{a}{2\sqrt{\Delta}}$. For small concentrations $\xi\gg a$, the state of the spin subsystem is close to the long-range order with the doubled elementary cell. Therefore, for $x=0.033$ the energies at ${\bf K}=(0,0)$ and $(\pi,\pi)$ are nearly degenerate. In Fig.~\ref{Fig1}, it is the only case with small Fermi surface pockets. With growing $x$ the correlation length approaches $a$, the dispersion acquires gradually the shape resembling a weakly correlated band. However, even for $x=0.14$ its bandwidth is much smaller than that for weak correlations. For the following consideration it is worth noting that Fermi surfaces are large for $x\gtrsim 0.06$.

\section{Results and discussion}
Below we consider the density of states,
\begin{equation}\label{dos}
\rho(\omega)=-\frac{1}{N\pi}\sum_{{\bf k}m}{\rm Im}G({\bf k}m\omega).
\end{equation}
For $T=0$ the integral of this quantity, multiplied by the frequency, over the occupied states gives the contribution of holes to the thermodynamic potential $\Omega$. As in the case of weak electron correlations, oscillations of $\rho(\omega=0)$ with the magnetic field lead to oscillations in $\Omega$ and its derivatives, which are observed in quantum oscillation measurements \cite{Shoenberg}. From (\ref{twopoles}) we get
\begin{equation}\label{rho}
\rho(\omega)=\frac{\phi}{N}\sum_{{\bf k}m}\frac{E_1({\bf k}m) -\varepsilon_-({\bf k}m)}{\varepsilon_+({\bf k}m) -\varepsilon_-({\bf k}m)} \delta[\omega-\varepsilon_-({\bf k}m)],
\end{equation}
where the dependencies of the energy parameters on ${\bf k}$ and $m$ are explicitly shown and it is taken into account that for the considered small hole concentrations only the lower pole of Green's function (\ref{twopoles}) crosses the Fermi level.

For several hole concentrations the frequency dependence of the DOS is shown in Fig.~\ref{Fig2}. As seen from this figure, $\rho(\omega)$ oscillates and these oscillations are connected with the applied magnetic field. Near the Fermi level the oscillations are not observed for $x=0.076$ and $x=0.18$. This absence of oscillations is connected with the sharp maxima of the DOS at the Fermi level. The oscillations are lost against the background of these maxima. The maxima of the DOS are pinned to the Fermi level in some ranges of concentrations below $x=0.076$ and around $x=0.18$, where the frequency oscillations near $\omega=0$ are also suppressed. Hence near the Fermi level the frequency oscillations of the DOS are observed in the range of hole concentrations $0.08<x<0.18$. The oscillations in the frequency dependence lead to oscillations in the dependence of $\rho(0)$ on the magnetic field. Notice that the indicated range of concentrations contains the interval, in which decreased quantum oscillation frequencies are observed experimentally \cite{Sebastian12}.

\begin{figure}
\resizebox{\textwidth}{!}{\includegraphics[width=\textwidth]{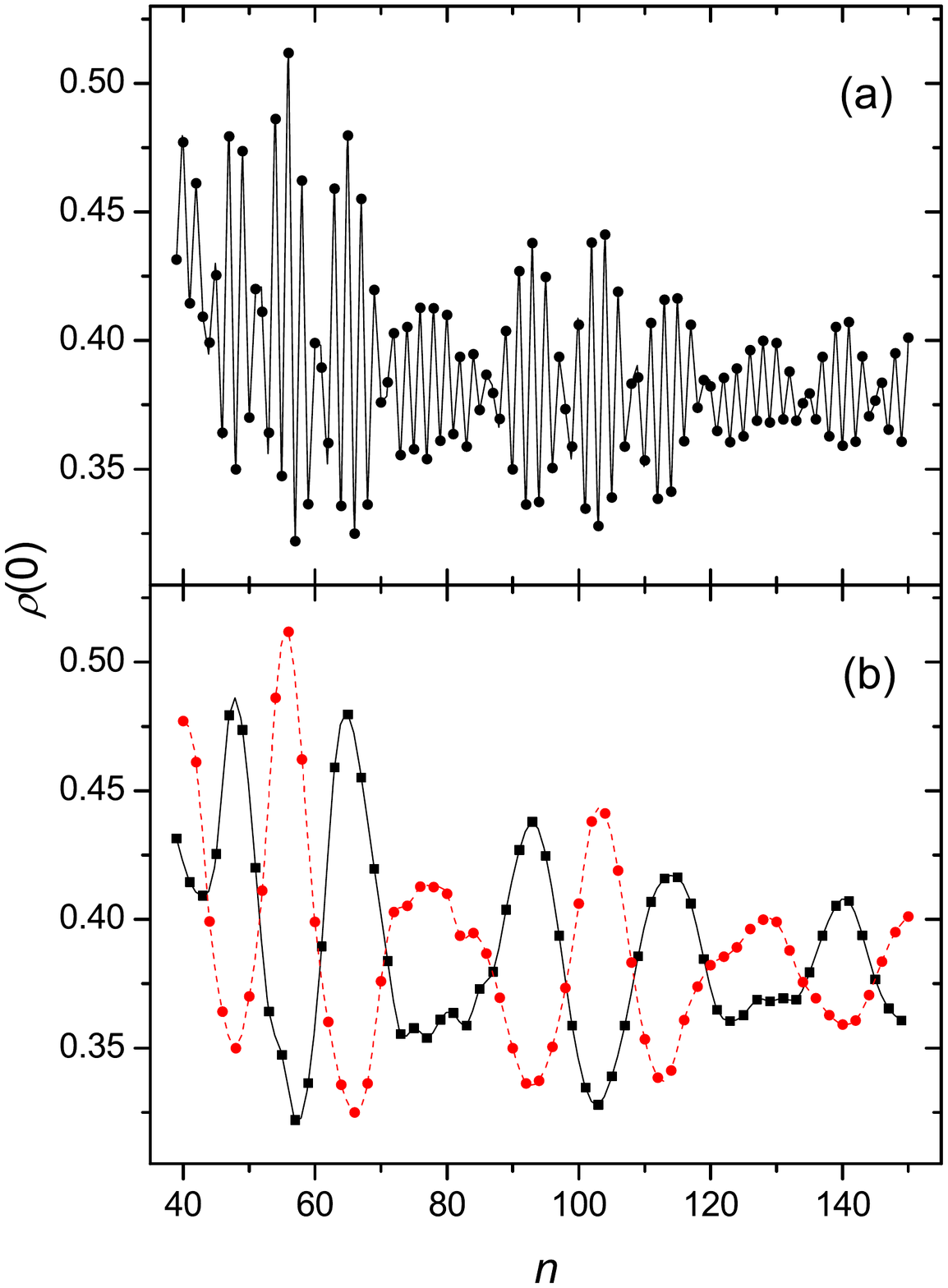}}
\caption{(a,b) The density of hole states at the Fermi level as a function of $n=\frac{2\pi\hbar}{ea^2}\frac{1}{B}$ for $x=0.14$. Calculated values are shown by symbols, connecting lines are splines. (a) $\rho(0)$ for all considered values of $n$, (b) $\rho(0)$ for even (red circles and red dashed line) and odd (black squares and black solid line) $n$. (c) The absolute value of the Fourier transform of the dependence $\rho(0)$ on $\frac{1}{B}$ for odd $n$.} \label{Fig3}
\end{figure}
The density of hole states at the Fermi level as a function of $n=\frac{2\pi\hbar}{ea^2}\frac{1}{B}$ is shown in Fig.~\ref{Fig3}a for the case $x=0.14$. Along with high-frequency oscillations with the period $\delta n=2-3$, which point to a large Fermi surface in this case, a modulation with a period which is larger by an order of magnitude is also observed.

The low-frequency oscillations become even more evident if we plot the dependence of $\rho(0)$ on $n\propto\frac{1}{B}$ for only even or only odd $n$. These dependencies are shown in Fig.~\ref{Fig3}b. Unexpectedly they are nearly in antiphase up to comparatively large $n$ and weak fields. For some other considered hole concentrations in the range $0.08<x<0.18$ the modulations are less evident than in Fig.~\ref{Fig3}a, which points to their sensitivity to the details of the hole dispersion (for all these $x$ the zero-field dispersions look similar to that in Fig.~\ref{Fig1}c with somewhat different bandwidths). Nevertheless, the nearly antiphase oscillatory dependencies of the type shown in Fig.~\ref{Fig3}b are well seen for all these $x$. This means that the low-frequency oscillatory dependence of the DOS on $\frac{1}{B}$ is inherent in systems with strong electron correlations. However, the possibility of its observation may depend on details of the hole dispersion.

The absolute value of the Fourier transform of the dependence $\rho(0)$ on $\frac{1}{B}$ for odd values of $n$ (black symbols in Fig.~\ref{Fig3}b) is shown in Fig.~\ref{Fig3}c as a function of the quantum oscillation frequency $F$. For this figure we set $a=4$~\AA. The Fourier transform of the second dependence in Fig.~\ref{Fig3}b has similar shape. In both these dependencies, peaks with the frequency $F_d\approx 1$~kT dominate. This frequency is of the same order of magnitude as the dominant frequency of quantum oscillations in underdoped YBa$_2$Cu$_3$O$_{6+x}$ \cite{Sebastian12}, which was interpreted as a manifestation of small Fermi surface pockets. Let us underline once again that the Fermi surface in Fig.~\ref{Fig1}c has no such pockets. As in the mentioned experiments, the dominant peak in Fig.~\ref{Fig3}c is flanked by weaker maxima.

\begin{figure}
\centerline{\resizebox{0.8\columnwidth}{!}{\includegraphics{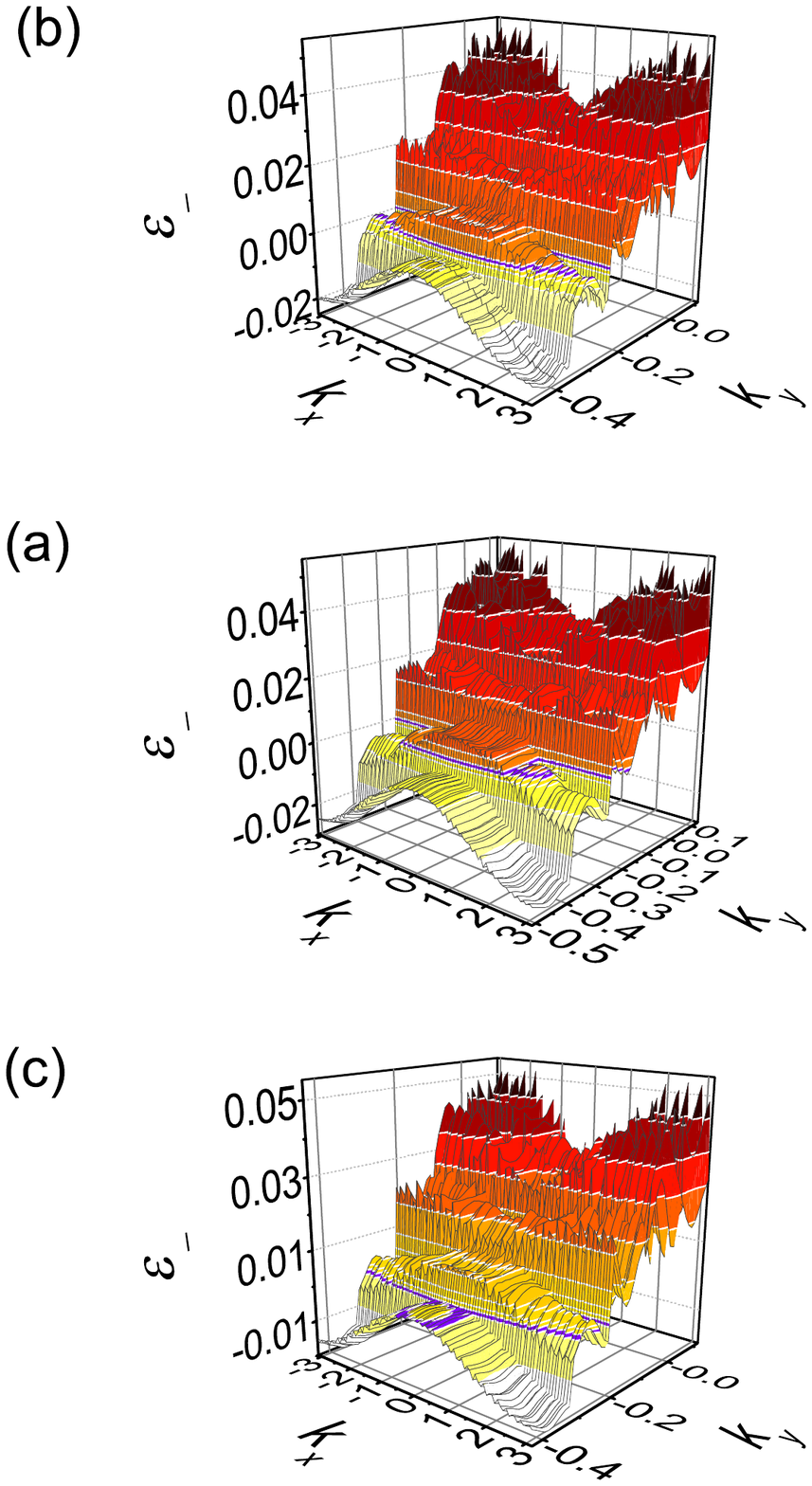}}}
\caption{The dispersion of hole states near the Fermi level at $x=0.14$ for $n=48$ (a), 52 (b) and 56 (c). The Fermi surface is shown by the violet (dark) contour.}\label{Fig4}
\end{figure}
What is the reason for the appearance of the low-frequency modulation in Fig.~\ref{Fig3}a and oscillations in \linebreak[4] Fig.~\ref{Fig3}b? In the uncorrelated case the carrier energy in the Landau subbands oscillates around some level when $k_x$ varies from $-\pi$ to $\pi$ at a fixed $k_y$. The number of oscillations is equal to $n$. Strong correlations lead to a bend of the subbands along the $x$ direction. This bending is seen in Fig.~\ref{Fig4}. In this figure, only the subbands near the Fermi level are shown. The subband $m$ is plotted in the stripe $-\pi+\frac{2\pi}{n}m\leq K_y<-\pi+\frac{2\pi}{n}(m+1)$, $-\pi\leq K_x<\pi$ of the entire Brillouin zone. As a result the hole dispersion looks like stairs with steps in the $y$ direction. These stairs ascend from smaller hole energies to larger ones with $K_y$ moving from $-\pi$ to $\pi$. A small part of these steps is shown in Fig.~\ref{Fig4}. The small-scale oscillations along the $x$ direction have the same origin as in the uncorrelated case. Bends of the subbands lead to the appearance of van Hove singularities, which produce maxima in the DOS. In Fig.~\ref{Fig4}c the chemical potential falls onto such a singularity and the curves in Figs.~\ref{Fig3}a and \ref{Fig3}b peak at $n=56$. Figure~\ref{Fig4}b ($n=52$) corresponds to some intermediate value of $\rho(0)$, while the Fermi level in Fig.~\ref{Fig4}a is halfway between two singularities, which leads to local minima in Fig.~\ref{Fig3} at $n=48$. Thus, the low-frequency modulation and oscillations in Fig.~\ref{Fig3} are connected with the traverse of the Fermi level through the sequence of van Hove singularities in the Landau subbands.

\section{Concluding remarks}
In this work, we have considered the two-dimensional $t$-$J$ model of Cu-O planes of cuprate perovskites under the conditions of strong electron correlations: $t\gg J$ and small hole concentrations $x$. The two-dimensional crystal is placed in a magnetic field, which is perpendicular to the crystal plane. Using the Mori projection operator technique we have calculated the Landau subbands in the case when only the nearest neighbor hopping constant is nonzero and for the magnetic field induction satisfying the condition $B=\frac{2\pi\hbar}{ea^2n}$, where $a$ is the lattice spacing and $n$ an integer. In the range of hole concentrations $0.08<x<0.18$ the density of hole states at the Fermi level $\rho(0)$ shows oscillations as a function of $n\propto\frac{1}{B}$. For somewhat smaller and larger hole concentrations the Fermi level falls onto strong maxima of the density of states, which hide the oscillations. In the range $0.08<x<0.18$, in the absence of the magnetic field the hole dispersion is characterized by large Fermi surfaces with areas, which are comparable to the area of the Brillouin zone. The period of the observed oscillation conforms with these large Fermi surface areas. However, in addition to the high-frequency oscillations the dependence $\rho(0)$ on $\frac{1}{B}$ shows a modulation with a frequency which is smaller by an order of magnitude. Considering $\rho(0)$ for even and odd $n$ we have obtained two low-frequency oscillatory dependencies, which are nearly in antiphase to each other up to comparatively large $n$ (small $B$). These nearly antiphase dependencies are observed for several considered hole concentrations in the range $0.08<x<0.18$.
The origin of these low-frequency modulation and oscillations is the bending of the Landau subbands near the Fermi level. The bending is a result of strong electron correlations. It leads to the appearance of van Hove singularities in the Landau subbands. For some values of $B$ the Fermi level falls onto these singularities, what produces the mentioned modulation of $\rho(0)$. The Fourier transformation of the dependencies $\rho(0)$ on $\frac{1}{B}$ for even and odd $n$ reveal the dominant frequency component $F_d\approx 1$~kT at $x=0.14$. This value is of the same order of magnitude as the dominant quantum oscillation frequency observed in underdoped YBa$_2$Cu$_3$O$_{6+x}$. It should be underlined that the low-frequency modulation and oscillations are not connected with small Fermi surface pockets -- in the considered range of $x$ the Fermi surface is large. Although the zero-field dispersions for these hole concentrations are similar, the low-frequency modulation may essentially differ. This points to the sensitivity of the modulations to details of the hole dispersion. In this context it is of interest to investigate the dependence of $\rho(0)$ on $B$ for fractional $n$ and for a more complicated and realistic zero-field dispersion. These points will be considered in the future.

\begin{acknowledgement}
This work was supported by the European Regional Development Fund (project TK114) and by the Estonian Scientific Foundation (grant ETF9371).
\end{acknowledgement}


\begin{thebibliography}{}
\bibitem{Castello}Castillo, H.E., Balseiro, C.A.: Phys.\ Rev.\ Lett.\ {\bf 68}, 121 (1992)
\bibitem{Beran}B\'eran, P.: Phys.\ Rev.\ B {\bf 54}, 1391 (1996)
\bibitem{Albuquerque}Albuquerque, A.F., Martins, G.B.; J.\ Phys.: Condens.\ Matter {\bf 17}, 2419 (2005)
\bibitem{Peierls}Peierls, R.: Z.\ Phys. {\bf 80}, 763 (1933)
\bibitem{Brown}Brown, E.: Phys.\ Rev.\ {\bf 133}, A1038 (1964)
\bibitem{Tripathi}Tripathi, G.S.: Phys.\ Rev.\ B {\bf 52}, 6522 (1995)
\bibitem{Yang}Yong Yang, MacDonald, A.H.: Phys.\ Rev.\ B {\bf 52}, R3876 (1995)
\bibitem{Schmid}Schmid, M., Andersen, B.M., Kampf, A.P., Hirschfeld, P.J.: New J.\ Phys.\ {\bf 12}, 053043 (2010)
\bibitem{Doiron}Doiron-Leyraud, N., Proust, C., LeBoeuf, D., Levallois, J., Bonnemaison, J.-B., Ruixing Liang, Bonn, D.A., Hardy, W.N., Taillefer, L.: Nature {\bf 447}, 565 (2007)
\bibitem{Bangura}Bangura, A.F., Goddard, P.A., Singleton, J., Tozer, S.W.,  Coldea, A.I., Ardavan, A., McDonald, R.D., Blundell, S.J., Schlueter, J.A.: Phys.\ Rev.\ B {\bf 76}, 052510 (2007)
\bibitem{Yelland}Yelland, E.A., Singleton, J., Mielke, C.H., Harrison, N.,  Balakirev, F.F., Dabrowski, B., Cooper, J.R.: Phys.\ Rev.\ Lett.\ {\bf 100}, 047003 (2008)
\bibitem{Sebastian08}Sebastian, S.E., Harrison, N., Palm, E., Murphy, T.P., Mielke, C.H., Ruixing Liang, Bonn, D.A., Hardy, W.N., Lonzarich, G.G.: Nature {\bf 454}, 200 (2008)
\bibitem{Shoenberg}Shoenberg, D.: Magnetic oscillations in metals. Cambridge University Press, Cambridge (1984)
\bibitem{Sebastian12}Sebastian, S.E., Harrison, N., Lonzarich, G.G.: Rep.\ Prog.\ Phys.\ {\bf 75}, 102501 (2012)
\bibitem{Millis}Millis, A. J., Norman, M.: Phys.\ Rev.\ B {\bf 76}, 220503(R) (2007)
\bibitem{Chen}Chen, W.-Q., Yang, K.-Y., Rice, T.M., Zhang, F.C.: EPL {\bf 82}, 17004 (2008)
\bibitem{Galitski}Galitski, V., Sachdev, S.: Phys.\ Rev.\ B {\bf 79}, 134512 (2009)
\bibitem{Melikyan}Melikyan, A., Vafek, O.: Phys.\ Rev.\ B {\bf 78}, 020502(R) (2008)
\bibitem{Pereg}Pereg-Barnea, T., Weber, H., Rafael, G., Franz, M.: Nature Phys.\ {\bf 6}, 44 (2010)
\bibitem{Varma}Varma, C.M.: Phys.\ Rev.\ B {\bf 79}, 085110 (2009)
\bibitem{Langbein}Langbein, D.: Phys.\ Rev.\ {\bf 180}, 633 (1969)
\bibitem{Hsu}Hsu, W.Y., Falicov, L.M.: Phys.\ Rev.\ B {\bf 13}, 1595 (1976)
\bibitem{Mori}Mori, H.: Progr.\ Theor.\ Phys. {\bf 34}, 399 (1965)
\bibitem{Sherman97}Sherman, A., Schreiber, M.: Phys.\ Rev.\ B {\bf 55},  R712 (1997)
\bibitem{Sherman06}Sherman, A.: Phys.\ Rev.\ B {\bf 74}, 035104 (2006)
\bibitem{Macridin}Macridin, A., Jarrell, M., Maier, T., Kent, P.R.C., D'Azevedo, E.: Phys.\ Rev.\ Lett.\ {\bf 97}, 036401 (2006)
\bibitem{Atkinson}Atkinson, W.A., Sonier, J.E.: Phys.\ Rev.\ B {\bf 77}, 024514 (2008)
\bibitem{Sherman87}Sherman, A.V.: J.\ Phys.\ A {\bf 20}, 569 (1987)
\bibitem{Sherman02}Sherman, A., Schreiber, M.: Phys.\ Rev.\ B {\bf 65}, 134520 (2002)
\bibitem{Sherman04}Sherman, A.: Phys.\ Rev.\ B {\bf 70}, 184512 (2004)
\bibitem{Izyumov}Izyumov, Yu.A.: Physics-Uspekhi {\bf 40}, 445 (1997)
\bibitem{Dagotto}Dagotto, E., Rev.\ Mod.\ Phys.\ {\bf 66}, 763 (1994)
\end{thebibliography}
\end{document}